\documentstyle[aps,multicol,epsfig]{revtex}

\newcommand{\be}{\begin{equation}}
\newcommand{\ee}{\end{equation}}

\begin{document}
\draft
\widetext

\title{Phase diagram of a Disordered Boson Hubbard Model in Two Dimensions\\} 
\author{Ji-Woo Lee$^{1}$, Min-Chul Cha$^{2}$ and Doochul Kim$^{1}$} 
\address {$^{1}$School of Physics and Center for Theoretical Physics, 
Seoul National University\\
 Seoul 151-742, Korea\\ 
$^{2}$Department of Physics, Hanyang University, Kyunggi-do, Ansan 425-791,
 Korea}

\maketitle

\thispagestyle{empty}

\begin{abstract}
We study the zero-temperature phase transition of a two-dimensional
disordered boson Hubbard model.
The phase diagram of this model is constructed in terms of the
disorder strength and the chemical potential.
Via quantum Monte Carlo simulations, we find a multicritical line
separating the weak-disorder regime, where a random potential is irrelevant,
from the strong-disorder regime.
In the weak-disorder regime, the Mott-insulator-to-superfluid transition
occurs, while, in the strong-disorder regime,
the Bose-glass-to-superfluid transition occurs.
On the multicritical line, the insulator-to-superfluid transition has
the dynamical critical exponent $z=1.35 \pm 0.05$ and the correlation length
critical exponent $\nu=0.67 \pm 0.03$,
that are different from the values for the transitions off the line.
We suggest that the proliferation of the particle-hole pairs screens out the
weak disorder effects.

\end{abstract}

          
\begin{multicols}{2}
\narrowtext 

The boson localization due to a random potential has continuously
attracted significant attention as a paradigm of the zero-temperature quantum
phase transition\cite{Sondhi97}.
The superconductor-insulator transition has been
believed to be a manifestation of the boson localization transition.
Two-dimensional realization of this transition may be found in  
disordered thin-film superconductors\cite{Markovic99},
Josephson-junction arrays\cite{Zant96},
and $^4$He films adsorbed in porous media\cite{Crowell95a}.
An interacting boson model, called boson Hubbard model,
has been proposed\cite{Fisher89}
to describe the transition between an insulator and a superfluid (SF).
In the disorder-free case, the insulating ground state is a Mott insulator (MI),
which has commensurate boson density and a finite Mott gap which suppresses
the excitations of freely moving particles or holes.
As the gap vanishes the system becomes a superfluid.
In the presence of disorder, even with vanishing energy gap,
particles or holes excited can be localized by a
random potential, resulting in a Bose glass (BG) insulator.

The interplay of interaction and disorder has attracted considerable
interest in this model.
It has been argued that in the presence of disorder
the transition from the insulating to the superfluid state occurs
only through the BG phase on the assumption that arbitrarily weak disorder
is always relevant in two dimensions\cite{Fisher89}.
Recently, however, quantum Monte Carlo studies of the boson Hubbard
model in the Villain representation have shown that
transitions at\cite{Kisker97} or near\cite{Park99} the tip of the MI lobes,
the points with the particle-hole symmetry,
are most likely direct from the MI state to the SF state for weak disorder.
Path-integral quantum Monte Carlo simulations\cite{Krauth91}
and real-space renormalization calculations\cite{Zhang92,Singh92}
of the boson Hubbard model
support the scenario of the direct MI-SF transition at the tips.
A simple scaling argument combined with renormalization calculation
at the mean-field level predicts\cite{Pazmandi98}
that the direct MI-SF is possible around the tip of the MI lobe
in high dimensions ($ d >4$)
while it is possible only at the tip in low dimensions ($2\leq d < 4$).
Field-theoretical renormalization group studies at the tip,
on the other hand, show that disorder is always relevant in
two dimensions\cite{Herbut98c,Kim94}.
Strong-expansion studies suggest that the direct
transition is always unlikely to occur in the presence of disorder\cite{Freericks96}.
Similar problems in one dimension have also attracted considerable interest
recently\cite{one-d}.
Here, it appears that a direct MI-SF transition is not supported.

\begin{figure}
\epsfxsize=9cm \epsfysize=6cm \epsfbox{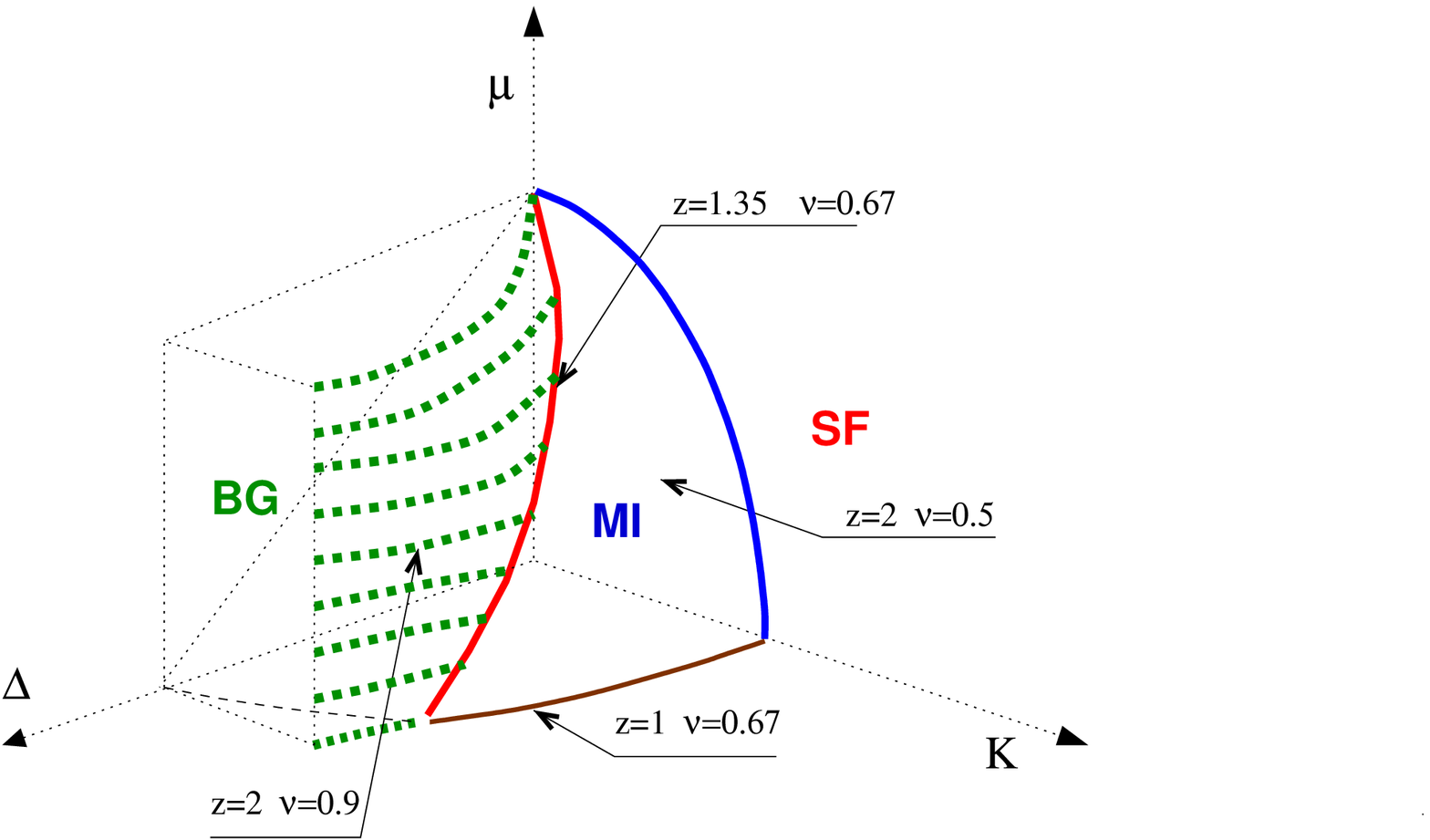}
\caption{Schematic phase diagram of a disordered boson Hubbard model.
Here $\mu$ is the chemical potential, $\Delta$ is the strength of disorder,
and $K$ is the parameter characterizing the hopping of bosons.
The particle-hole symmetry is defined by the condition $\mu=0$.
The multicritical line separates the critical surface into
the direct MI-SF transition region and the BG-SF transition region.}
\label{fig1}
\end{figure}

In this work, we investigate the onset of the superfluidity
in a two-dimensional boson Hubbard model
in the presence of disorder via quantum Monte Carlo simulations
which employ a (2+1)-dimensional classical action.
We find that a multicritical line exists, separating the weak-disorder regime
from the strong-disorder regime, and on the line,
the insulator-to-superfluid transition is associated with novel values of
the critical exponents: i.e. $z=1.35 \pm 0.05$ and $\nu=0.67 \pm 0.03$,
where $z$ and $\nu$ are the dynamical and
the correlation length critical exponents, respectively.
These results are summarized in Figure~\ref{fig1}.
It shows that the direct MI-SF transition survives around the tip of
the MI lobe for weak disorder.
This means that the weak disorder is irrelevant for the localization transition
in two dimensional interacting boson systems that have near integer fillings.
Strong disorder changes the nature of the transition to that of 
the BG-SF transition.

We consider a boson Hubbard model on the two-dimensional square lattice,
given by the Hamiltonian
\begin{equation}
H= \frac{1}{2} \sum_{i} U n_i^{2} - \sum_{i} \mu_i n_i
- \frac{t}{2} \sum_{\langle i,j \rangle}
(b_i b^{\dagger}_j + b^{\dagger}_i b_j) \ \ ,
\label{eq1}
\end{equation}
where $b_i^{\dagger}$ ($b_i$) denotes the boson creation (destruction)
operator at site $i$ ($n_i \equiv  b_i^{\dagger} b_i $),
$\mu_i$ the local chemical potentials,
$U$ the on-site repulsion energy,
$t$ the hopping strength to the nearest neighbors,
and finally the last sum is over nearest neighbor pairs.  
Disorder effect is embedded in local chemical potential as $\mu_i = \mu + v_i$
where $v_i$ are random variables independently and uniformly distributed
in the range $[-\Delta, \Delta]$.
Thus $\Delta$ characterizes the strength of disorder.

To perform Monte Carlo simulations,
we follow the standard procedure\cite{Wallin94}
to transform the two-dimensional Hamiltonian in Eq.~(\ref{eq1})
to the (2+1)-dimensional classical action 
\begin{equation}
S= \frac{1}{2K} \sum_{(x,y,\tau)}^{\nabla \cdot \vec{J}=0}
[\vec{J}^2 (x,y,\tau) - 2(\mu + v_i ) J^{\tau}(x, y, \tau)]\ \ ,
\label{action}
\end{equation}
where $x$ and $y$ are spatial coordinates and $\tau$ is imaginary-temporal
coordinate.
Here $K \sim \sqrt{t/U}$, analogous to the temperature in classical systems,
and $\vec{J}(x,y,\tau)$ is the integer current vectors which measure
the fluctuations along the corresponding direction of the components.
In the transformation from Eq.~(\ref{eq1}) to Eq.~(\ref{action}) one
assumes that superfluidity is destroyed only by phase fluctuations.

We are interested in the phase diagram of Eq.~(\ref{action}) in the
space of $(K,\mu,\Delta)$.
The transition described by the classical action is studied by
Monte Carlo simulations.
We perform the simulations on the lattice of various sizes, denoted by
$L \times L \times L_\tau$, where $L$ and $L_\tau$ are sizes of the systems
along a spatial and the imaginary-temporal axis respectively.
The periodic boundary conditions are adopted.
In order to extract the critical properties of the transition,
we analyze the data using the finite-size scaling theory.
An important quantity which indicate the onset of the superfluidity 
is the superfluid stiffness, which is measured by the formula
\begin{equation}
\rho = \frac{1}{L_\tau} [ \langle n_x^2 \rangle ]_{av}
\end{equation}
where $n_x = (1/L) \sum_{(x,y,\tau)} J^x(x,y, \tau)$ is the winding
number along the $x$ direction. Here $[...]_{av}$ denotes the average over
different realizations of disorder.
The finite-size scaling behavior of the superfluid stiffness
is given by\cite{Cha91}
\begin{equation}
\rho = L^{-(d+z-2)} \tilde{\rho} ( L^{1/\nu} \delta, L_\tau / L^z )\ \ ,
\label{rho_scale}
\end{equation}
where $\delta = (K-K_c)$ is the distance from the critical point $K_c$
for each $\Delta$ and $\mu$,
$\tilde{\rho}$ is a scaling function,
and $d$ is the spatial dimension.
Throughout this work, we set $d=2$.
Another useful quantity is the compressibility, which directly shows
whether the Mott gap vanishes at the transition.
It is given by the formula
\begin{equation}
\kappa = \frac{L_\tau}{L^d} [ \langle {n_\tau}^2 \rangle
- \langle {n_\tau} \rangle^2]_{av}
\end{equation}
with $n_\tau = (1/L_\tau) \sum_{(x,y,\tau)} J^\tau (x,y, \tau)$,
and is assumed to take the scaling form
\begin{equation}
\kappa = L^{z-d} \tilde{\kappa} ( L^{1/\nu} \delta, L_\tau / L^z ) \ \ .
\end{equation}
with another scaling function $\tilde{\kappa}$.

In order to investigate the scaling behavior of the superfluid stiffness
and the compressibility,
one must specify the dynamical critical exponent, $z$, in advance
so as to fix the aspect ratio $L_\tau / L^z$ throughout the simulations.
We have tried various values of $z$ and have chosen the one which gives
the best scaling behavior satisfying Eq.~(\ref{rho_scale}).
For lattices with non-integer $L_\tau$, the superfluid stiffness is
obtained by a simple interpolation of the two values measured in the lattices
of nearby integer sizes.

The effect of disorder at various strength is investigated on $\mu=0$ plane.
For strong disorder, say for $\Delta > 0.45$, the onset of superfluidity
follows the BG-SF transition behavior with $z=2$ and $\nu=0.9$,
and the compressibility is finite at the transition as suggested
by the scaling argument\cite{Fisher89}
and confirmed by subsequent numerical simulations\cite{Wallin94}.
For $\Delta <0.35$, however, the transitions take the signature
of the direct MI-SF transition with $z=1$ and $\nu=0.67$,
as reported in the recent simulations\cite{Kisker97},
while the compressibility vanishes at the transition.
At the intermediate strength of disorder, 
naive scaling analysis using either $z=1$ or $z=2$ fails.
In this work, we take a different approach and allow the possibility
of intermediate values of $z$ by varying the aspect ratio $L_\tau / L^z$
until the scaling plots collapse onto a single curve.

A new scaling behavior emerges, at the intermediate strength of disorder,
which has $z=1.35 \pm 0.05$ and $\nu=0.67 \pm 0.03$.
This scaling behavior is shown in Figure~\ref{fig2}.
We observe the best scaling when $z=1.35$.
In addition, with this value of $z$, $K_c$'s obtained from the scaling behavior
of the superfluid stiffness and the compressibility data curves
are consistent with each other.
The compressibility vanishes at the transition as expected
for $z < d$ (see Figure~\ref{fig2}(b)).
Kisker and Rieger reported\cite{Kisker97} $z \approx 1.4$ for the intermediate
strength of disorder, $\Delta=0.4$.
However, they interpreted the result as simply measuring an effective exponent.
As we can see in Figure~\ref{fig2}, the robust finite-size scaling behavior
for various sizes strongly suggests
that the transition truly has new values of the critical exponents.
In order to rule out the possibility that we are disguised 
by finite size effects,
we have performed simulations adopting a different aspect ratio
and find the same critical exponents. 
The value of the correlation length critical exponent 
happens to be the same as the pure transition at the tip.

This behavior persists even off the tip ($\mu \ne 0$).
The off-tip transition of the pure case, usually called the generic transition,
is of a mean-field type,
so that $z=2$ and $\nu=1/2$ for the MI-SF transition.
If disorder is relevant, we expect in general
that the correlation length critical exponent have a different value\cite{nu}.
Thus, we can numerically identify the BG-SF transition from the
MI-SF transition by measuring $\nu$ as well as by observing the compressibility.
Previously Park {\it et~al.} have reported\cite{Park99}
that even in the presence of weak disorder,
the generic transition survives near the tip.
As $\Delta$ increases, the generic MI-SF transition ends at
the critical value, $\Delta_c(\mu)$, above which the BG-SF transition occurs.

The critical strength of disorder, $\Delta_c(\mu)$, therefore,
defines a multicritical line,
separating the critical surface $K_c(\mu,\Delta)$ into the two regions:
the strong-disorder region and the weak-disorder one.
Figure~\ref{fig3} is the measured $\Delta_c(\mu)$.
Numerically we find good scaling behaviors using the above exponents 
for a range of parameters (see the inset of Figure~\ref{fig3}). 
The error bars in Figure~\ref{fig3} show such ranges.
We believe that, in larger systems, the finite range should shrink to a point.
Otherwise, the existence of an incompressible localized phase other than
the BG phase is required, which is unreasonable.

The existence of the multicriticality strongly suggests that
in the presence of disorder two different universality classes exist,
in one of which disorder is irrelevant.
The irrelevance of weak disorder may be understood from the abundance of the
particle-hole pairs excited.
At the tips, the proliferation of particle-hole pairs near the transition
modifies the particle propagation to yield $z=1$ (otherwise
$z=2$) in the pure case.
These particle-hole pairs may screen the disorder effects.
In order to directly check the possibility that the proliferation of
particle-hole pairs may screen out the weak disorder effects,
we have repeated Monte Carlo simulations of the classical model given by
Eq.~(\ref{action}) under the condition
that the background particle-hole fluctuations are completely suppressed.
For simulations of the system governed by the action of Eq.~(\ref{action}),
one should generate all possible current configurations
$\{{\vec J}(x,y,\tau)\}$\cite{Wallin94}.
The suppression of the particle-hole fluctuations is realized
by disallowing any Monte Carlo updating which
generates local particle-hole current loops.
In the spatial planes, however, all possible current configurations are
allowed as usual.
Thus the allowed current configurations,
if they are not restricted in a spatial plane,
percolate in the $\tau$-direction (the imaginary temporal direction).

Figure~\ref{fig4} shows the scaling behavior of the superfluid stiffness
when the particle-hole fluctuations are suppressed at the tip ($\mu=0$)
for relatively weak disorder ($\Delta=0.3$).
The scaling shows that the critical exponents are $z=2$ and $\nu=0.9$,
indicating the BG-SF transition occurs.
Note that, with the particle-hole fluctuations, the transition
has the behavior of the MI-SF transition\cite{Kisker97,Park99}
at this strength of disorder.
This simulation result, therefore, are consistent with the argument that
the particle-hole pair fluctuations make the weak disorder irrelevant to
the transition.

Now one may question how the particle-hole fluctuations screen the disorder.
In the mean-field picture, ignoring the particle-hole fluctuations,
particles (holes) are localized by arbitrarily weak disorder due to
the coherent impurity back-scattering.
However, the proliferation of the particle-hole pairs could hinder the coherent
back-scattering. A particle propagating 
could be captured and destroyed by holes, and a new particle starts to
propagate. A phase shift
which accompanies such particle-exchange process may destroy
the coherent back-scattering of the particle.
As a result, this effect could screen out a weak random potential,
making it irrelevant.
Slightly off the tip, there are still abundant particle-hole pairs fluctuating.
Thus this screening effect will survive even off the tip.

In summary,
via quantum Monte Carlo simulations of the phase-only model of a disordered
boson Hubbard model,
we have found that there is a multicritical line which divides
the universality of this model into
the weak-disorder regime and the strong-disorder regime.
In the weak-disorder regime, disorder is irrelevant for the
insulator-to-superfluid transition.
On the multicritical line, the dynamical critical exponent $z=1.35 \pm 0.05$ 
and the correlation exponent $\nu=0.67 \pm 0.03$ are obtained.
Our result strongly suggests the direct MI-SF transition survives
for weak disorder.
We argue that the particle-hole pairs fluctuating near the tip will
screen out disorder to make a weak random potential irrelevant.

\acknowledgements

The work of M.C.C. has been supported by Korea Research Foundation
Grant(KRF2000-041-D00125).

\begin{figure}
\vspace{3.5cm}
\epsfxsize=6cm \epsfysize=6cm \epsfbox{fig2a.eps}
\end{figure}
\begin{figure}
\vspace{2cm}
\epsfxsize=6cm \epsfysize=6cm \epsfbox{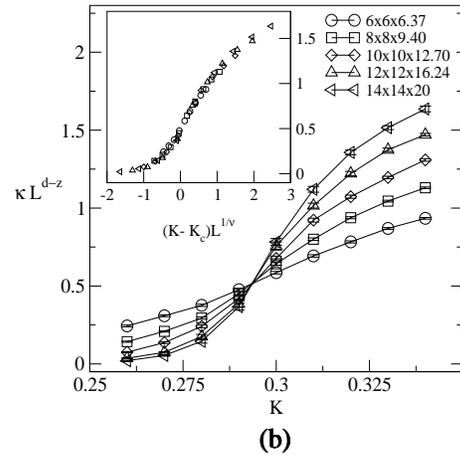}
\caption{The finite size scaling behavior of (a) the superfluid stiffness
and (b) the compressibility
at the tip of the MI lobe ($\mu=0$) with $\Delta=0.40$.  We set $z=1.35$.
The curves cross at $K_c=0.292$. The inset shows the
scaled data along $K$-direction with $\nu=0.67$.}
\label{fig2}
\end{figure}

\begin{figure}
\vspace{3.5cm}
\epsfxsize=6cm \epsfysize=6cm \epsfbox{fig3.eps}
\caption{The critical strength of disorder, $\Delta_c (\mu)$,
as a function of the chemical potential $\mu$.
At the tip ($\mu=0$), we have tuned the value of $\Delta$ to find
the critical one. The error bar indicates the range in which
the scaling plots show the behavior expected on the multicritical point.
Off the tip $\mu$ is tuned instead, while $\Delta$ is fixed.
Inset: The superfluid stiffness scaling curves away from the tip ($\mu=0.175$)
at $\Delta=0.2$ with $z=1.35$ and $\nu=0.67$.}
\label{fig3}
\end{figure}

\begin{figure}
\vspace{2.5cm}
\epsfxsize=6cm \epsfysize=6cm \epsfbox{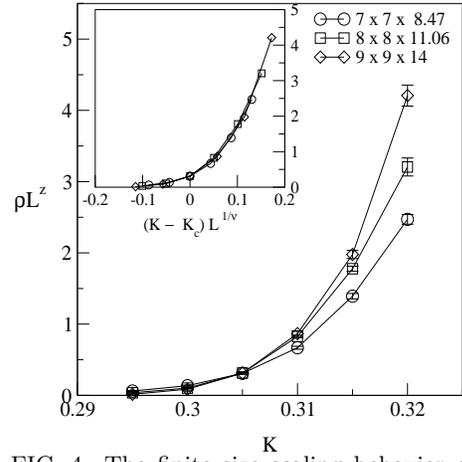}
\caption{The finite size scaling behavior of the superfluid stiffness
at $\Delta=0.3$ ($\mu=0$) for the systems in which the particle-hole
pair fluctuations are completely suppressed.
We set $z=2$.  The inset shows the
scaled data along $K$-direction with $\nu=0.9$.
These exponents indicate that the nature of the transition is of the BG-SF one.}
\label{fig4}
\end{figure}

\end{multicols}
\end{document}